\documentclass[conference]{IEEEtran}
\IEEEoverridecommandlockouts

\usepackage{cite}
\usepackage{amsmath,amssymb,amsfonts}
\usepackage{algorithmic}
\usepackage{graphicx}
\usepackage{textcomp}
\usepackage{xcolor}
\def\BibTeX{{\rm B\kern-.05em{\sc i\kern-.025em b}\kern-.08em
    T\kern-.1667em\lower.7ex\hbox{E}\kern-.125emX}}

\usepackage{enumitem}
\usepackage{amsthm}
\usepackage{multirow}
\usepackage{subcaption}
\usepackage{xspace}
\usepackage{booktabs}
\usepackage{lipsum} 
\usepackage{algorithm, algorithmic}
\usepackage{threeparttable}
\usepackage{url}
\usepackage{array}
\newcolumntype{C}[1]{>{\centering\arraybackslash}p{#1}}

\usepackage{makecell}

\newtheorem{proposition}{Proposition}

\newcommand{\method}[0]{SCAR\xspace}
\newcommand{\methodlong}[0]{Simple Collaborative Augmentation for Recommendation\xspace}    
\newcommand{\moduleadd}[0]{\textsc{ColAdd}\xspace}
\newcommand{\moduleaddlong}[0]{Collaborative Edge Addition\xspace}
\newcommand{\modulerep}[0]{\textsc{ColRep}\xspace}
\newcommand{\modulereplong}[0]{Collaborative Edge Replacement\xspace}

\begin{document}

\title{Simple and Behavior-Driven Augmentation for Recommendation with Rich Collaborative Signals}


\author{\IEEEauthorblockN{Doyun Choi}
\IEEEauthorblockA{School of Electrical Engineering \\
KAIST\\
Daejeon, South Korea \\
doyun.choi@kaist.ac.kr}
\and
\IEEEauthorblockN{Cheonwoo Lee}
\IEEEauthorblockA{School of Electrical Engineering \\
KAIST\\
Daejeon, South Korea \\
cheonwoo.lee@kaist.ac.kr}
\and
\IEEEauthorblockN{Jaemin Yoo}
\IEEEauthorblockA{School of Electrical Engineering \\
KAIST\\
Daejeon, South Korea \\
jaemin@kaist.ac.kr}
}

\maketitle

\begin{abstract}
Contrastive learning (CL) has been widely used for enhancing the performance of graph collaborative filtering (GCF) for personalized recommendation.
Since data augmentation plays a crucial role in the success of CL, previous works have designed augmentation methods to remove noisy interactions between users and items in order to generate effective augmented views.
However, the ambiguity in defining ``noisiness'' presents a persistent risk of losing core information and generating unreliable data views, while increasing the overall complexity of augmentation. 
In this paper, we propose \methodlong (\method), a novel and intuitive augmentation method designed to maximize the effectiveness of CL for GCF.
Instead of removing information, SCAR leverages collaborative signals extracted from user-item interactions to generate pseudo-interactions, which are then either added to or used to replace existing interactions.
This results in more robust representations while avoiding the pitfalls of overly complex augmentation modules. 
We conduct experiments on four benchmark datasets and show that \method outperforms previous CL-based GCF methods as well as other state-of-the-art self-supervised learning approaches across key evaluation metrics. 
\method exhibits strong robustness across different hyperparameter settings and is particularly effective in sparse data scenarios.
Our implementation is available at \url{https://github.com/cdy9777/SCAR}.


\end{abstract}


\section{Introduction}


Recommender systems are essential tools for personalized content delivery on many platforms, such as social media \cite{guy2009you}, e-commerce \cite{schafer1999recommender}, and streaming services \cite{1423975}.
As the amount of available content continues to surge, effective recommender systems are crucial for helping users navigate extensive catalogs and discover content that matches their interests \cite{Ricci2022, 5197422}. 
Accurate recommendation not only enhances user experience but also improves business performance by predicting user preferences and offering relevant suggestions. 

\begin{figure}[t]
\centering
\includegraphics[width=\columnwidth]{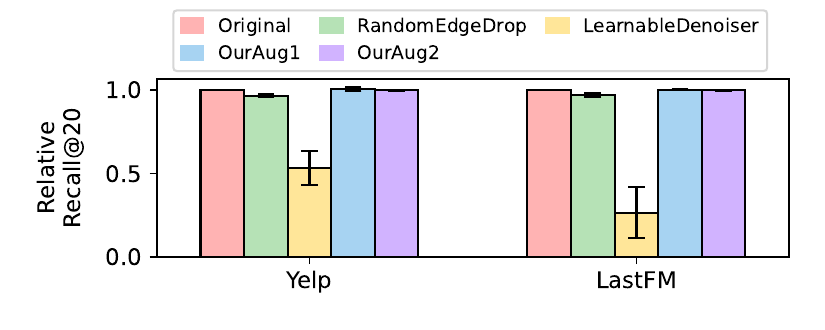}
\vspace{-25pt}
\caption{
    Performance comparison on the Yelp and LastFM datasets using the original graph for training and the augmented graphs for inference.
    ``Denoising'' original edges degrades performance, even with a learnable denoiser \cite{10.1145/3580305.3599768}, highlighting its risk.
    In contrast, augmented graphs from our methods, i.e., OurAug1 and OurAug2, remain stable. 
}
\label{fig:Validity of Denoising}
\vspace{-15pt}
\end{figure}

Collaborative filtering (CF) is a fundamental technique in recommender systems that leverages patterns in user-item interactions to generate personalized recommendations \cite{4781121,10.1145/3038912.3052569, sarwar2001item}.
By analyzing historical user behaviors such as ratings or clicks, CF predicts future interactions by identifying similar users or items.
Recently, graph neural networks (GNNs) have gained significant interest for analyzing dynamically structured data, leading to the development of graph-based collaborative filtering (GCF) \cite{10.1145/3331184.3331267}.
By aggregating information across multi-hop connections within a graph, GNNs provide richer representations of users and items, thereby improving the overall accuracy of recommendations \cite{9046288, kipf2016semi, NIPS2017_5dd9db5e}.
Notable models like NGCF \cite{10.1145/3331184.3331267} and LightGCN \cite{10.1145/3397271.3401063} have demonstrated exceptional performance in personalized recommendation tasks.

However, challenges such as data sparsity, long-tail distributions, and noisy interactions limit the performance of GCF by reducing available interactions and introducing bias.
To address these limitations, contrastive learning (CL) has emerged as a promising solution.
The key idea of CL is to help the model learn informative signals by contrasting different views of the data, generated from various augmentation functions.
CL enhances overall recommendation performance by improving both the uniformity and distinctiveness of the learned representations of items and users \cite{pmlr-v119-chen20j, pmlr-v119-wang20k}.


The effectiveness of CL largely depends on the quality of data augmentation functions used to generate contrastive views.
From this perspective, beyond random perturbations of interaction graphs \cite{10.1145/3404835.3462862}, recent approaches have aimed to generate augmented data by filtering out noisy interactions.
For example, CGI \cite{NEURIPS2022_803b9c4a} generates contrastive views using a learnable mask guided by an information bottleneck objective. AdaGCL \cite{10.1145/3580305.3599768} employs a parameterized edge-denoising network, inspired by PTDNet \cite{luo2020learningdroprobustgraph}, to generate contrastive views.
DCCF \cite{10.1145/3539618.3591665} creates noise-masked augmented graphs based on intent-aware pattern analysis.

A key limitation of recent augmentation methods lies in the difficulty of defining “noisy interactions” in recommendation datasets.
Unlike other domains, most recommender systems rely on implicit feedback \cite{10.1145/3038912.3052569, 4781121}, which only indicates the presence or absence of interactions. 
These interactions often result from a complex mix of latent factors—including user intent (e.g., curiosity, exploration), item popularity, context, or interface design—making it difficult to objectively identify and remove noise without risking the loss of meaningful signals.
Figure~\ref{fig:Validity of Denoising} shows the result of our toy experiment to intuitively demonstrate the risk of denoising.
Denoised graphs harm the performance of recommendation when used at inference, creating a bias in users' true intents (details in Section \ref{ssec:Motivating Experiment}).

In this paper, we adopt a different approach from existing works, aiming to reduce the risk of core interaction loss while providing meaningful information to the generated views.
We propose Simple Collaborative Augmentation for Recommendation (SCAR), a novel augmentation method that leverages the collaborative information inherent in user-item interactions.
The idea is to generate pseudo-interactions based on the collaborative signals and add them to the given data, rather than removing existing information.

We introduce two key augmentation functions to achieve this goal: collaborative edge addition (\moduleadd) and collaborative edge replacement (\modulerep).
In \moduleadd, we derive \emph{effectiveness scores} of items to users, and add pseudo-interactions that are expected to enrich the collaborative signals.
In \modulerep, we replace the least effective edges with the most similar ones based on collaborative item-item similarities.
These functions generate augmented views with diverse CF signals, enabling \method to improve user and item representations through effective contrastive learning while maintaining a simple, efficient, and interpretable framework.


In summary, our contributions are given as follows:
\begin{itemize}[topsep=5pt, leftmargin=10pt, itemsep=0.5em]

\item \textbf{Algorithms:} We propose \method, which generates diverse augmented views for recommendation while reducing the risk of losing core interactions.
\method consists of \moduleadd and \modulerep, augmentation functions for leveraging collaborative information to generate pseudo-interactions.

\item \textbf{Analysis:} We thoroughly analyze the core advantages of \method.
The augmented views generated by \method enable the encoder to capture multi-hop signals that are unreachable by trivial GNN encoders with a limited number of layers.
Additionally, we analyze the time complexity of \method and show its scalability.

\item \textbf{Experiments:} Through extensive experiments, we demonstrate that \method achieves superior performance over eight baseline methods across four datasets in most cases, while relying on simple and intuitive augmentation techniques.
\end{itemize}

\section{Preliminaries and Related Works}


\subsection{Problem Definition}


We solve a recommendation problem with implicit feedback, where user preferences are inferred from observed interactions with items rather than explicit ratings.
We denote the user set as $U=\{u_k\}_{k=1}^m$ and the item set as $I=\{i_k\}_{k=1}^n$, where $m$ is the number of users and $n$ is the number of items.
We are given as input the relation matrix $\mathbf{R} \in \mathbb{R}^{m \times n}$, which contains the observed interactions between users and items. For each element $r_{u,i}$, if an interaction between user $u$ and item $i$ is observed, then $r_{u,i} = 1$; otherwise, $r_{u,i} = 0$.
Our goal is to predict the test users' preferences for items by utilizing encoded representations of users and items derived from $\mathbf{R}$.

\subsection{Graph-based Collaborative Filtering}

Graph-based collaborative filtering (GCF) enhances user and item representations by capturing various collaborative signals using graph neural networks (GNNs), understanding the interaction data as a bipartite graph. 
Among various GNN encoders, we use LightGCN \cite{10.1145/3397271.3401063}, which effectively captures collaborative signals with reduced complexity \cite{10.1145/3580305.3599768, 10.1145/3485447.3512104, 10.1145/3477495.3531937}.


Given $\mathbf{R}$, we first construct a graph $G = (V, E)$ with the node set $V = U \cup I$ and the edge set $E = \{(u, i) \mid u \in U, i \in I, r_{u,i} = 1\}$.
This graph can be represented by the following adjacency matrix $\mathbf{A}$:
\begin{equation}
    \label{eq:Adjacency matrix}
\mathbf{A} = \begin{bmatrix} 
\mathbf{0} & \mathbf{R} \\
\mathbf{R}^\mathrm{T} & \mathbf{0}
\end{bmatrix}.
\end{equation}
We then normalize $\mathbf{A}$ into $\mathbf{\tilde{A}} = \mathbf{D}^{-\frac{1}{2}} \mathbf{A} \mathbf{D}^{-\frac{1}{2}}$, where $\mathbf{D}$ is the degree matrix of $\mathbf{A}$.
The initial node embedding matrix $\mathbf{E}^{(0)} \in \mathbb{R}^{(m+n) \times d}$ consists of learnable parameters, where $d$ represents the embedding size.
The LightGCN encoder propagates information through the graph for $L$ layers using the following update rule:
\begin{equation}
    \label{eq02}
\mathbf{E}^{(l)} = \mathbf{\tilde{A} E}^{(l-1)}, \quad l = 1, 2, \dots, L,
\end{equation}
where $\mathbf{E}^{(l)}$ represents the node embedding matrix at layer $l$. 
Note that $\mathbf{E}^{(l)}$ is composed of $\mathbf{E}^{(l)}_U$ and $\mathbf{E}^{(l)}_I$, which represent the embeddings for users and items, respectively.

After $L$ layers of propagation, we apply a readout function to obtain the final representations of nodes:
\begin{equation}
\label{eq03}
\mathbf{Z}_{U} = {\textstyle \sum_{l=0}^{L}} \mathbf{E}^{(l)}_U, \quad \mathbf{Z}_{I} = {\textstyle \sum_{l=0}^{L}} \mathbf{E}^{(l)}_I,
\end{equation}
where $\mathbf{Z}_U$ and $\mathbf{Z}_I$ represent the final embeddings for all users and items, respectively.
Using these representations, we predict the interaction between user $u$ and item $i$ as $\hat{y}_{u,i} = \mathbf{z}_u^{\mathrm{T}} \mathbf{z}_i$.


We train the model based on the Bayesian Personalized Ranking (BPR) loss \cite{rendle2012bprbayesianpersonalizedranking}, a common objective function for collaborative filtering.
The BPR loss is defined as
\begin{equation}
\textstyle \mathcal{L}_{\text{BPR}} = - \sum_{(u,i,j) \in \mathcal{D}} \log \sigma (\hat{y}_{u,i} - \hat{y}_{u,j}),
\end{equation}
where $\sigma$ is the sigmoid function, and $\mathcal{D}=\{(u, i, j) \mid r_{u,i}=1, r_{u,j}=0\}$ is the set of positive and negative samples.
This encourages the predicted score $\hat{y}_{u,i}$ for observed interactions to be higher than that for non-interacted ones $\hat{y}_{u,j}$.


\subsection{Contrastive Learning for Recommendation}

Contrastive learning (CL) has been widely applied across various domains to enhance the learning capability of encoders \cite{pmlr-v119-chen20j, pmlr-v119-wang20k}. 
Building on the success of CL in various fields \cite{gao2022simcsesimplecontrastivelearning, He_2020_CVPR}, previous works have used CL to improve the performance of GCF.
Notable models that integrate CL with GCF include SGL \cite{10.1145/3404835.3462862}, SimGCL \cite{10.1145/3477495.3531937}, HCCF \cite{10.1145/3477495.3532058}, and NCL \cite{10.1145/3485447.3512104}.
By incorporating a CL loss function, these approaches help the encoder capture more informative and discriminative features, resulting in improved recommendation performance.

Recently, numerous augmentation techniques have been developed to enhance the effectiveness of CL by reflecting properties such as diversity, validity, and domain-adaptivity.
In particular, methods focusing on validity aim to preserve the semantic information of nodes while filtering out noisy interactions, leading to increasingly complex strategies \cite{10.1145/3580305.3599768, NEURIPS2022_803b9c4a}.
However, defining noise remains ambiguous since interactions in recommendation data capture personal user preferences, making direct adoption of denoising techniques from other domains risky \cite{vignac2023digressdiscretedenoisingdiffusion, 10.1145/3437963.3441734}. These challenges are compounded by higher computational costs and reduced interpretability, limiting the practical effectiveness of such approaches.

In response to these challenges, we introduce an effective augmentation method for GCF.
It provides rich collaborative signals while preserving core interactions, allowing the encoder to effectively capture the key features.
Our approach also prioritizes simplicity and interpretability, ensuring performance improvement without adding unnecessary complexity.


\begin{figure}
    \centering
    \includegraphics[width=\linewidth]{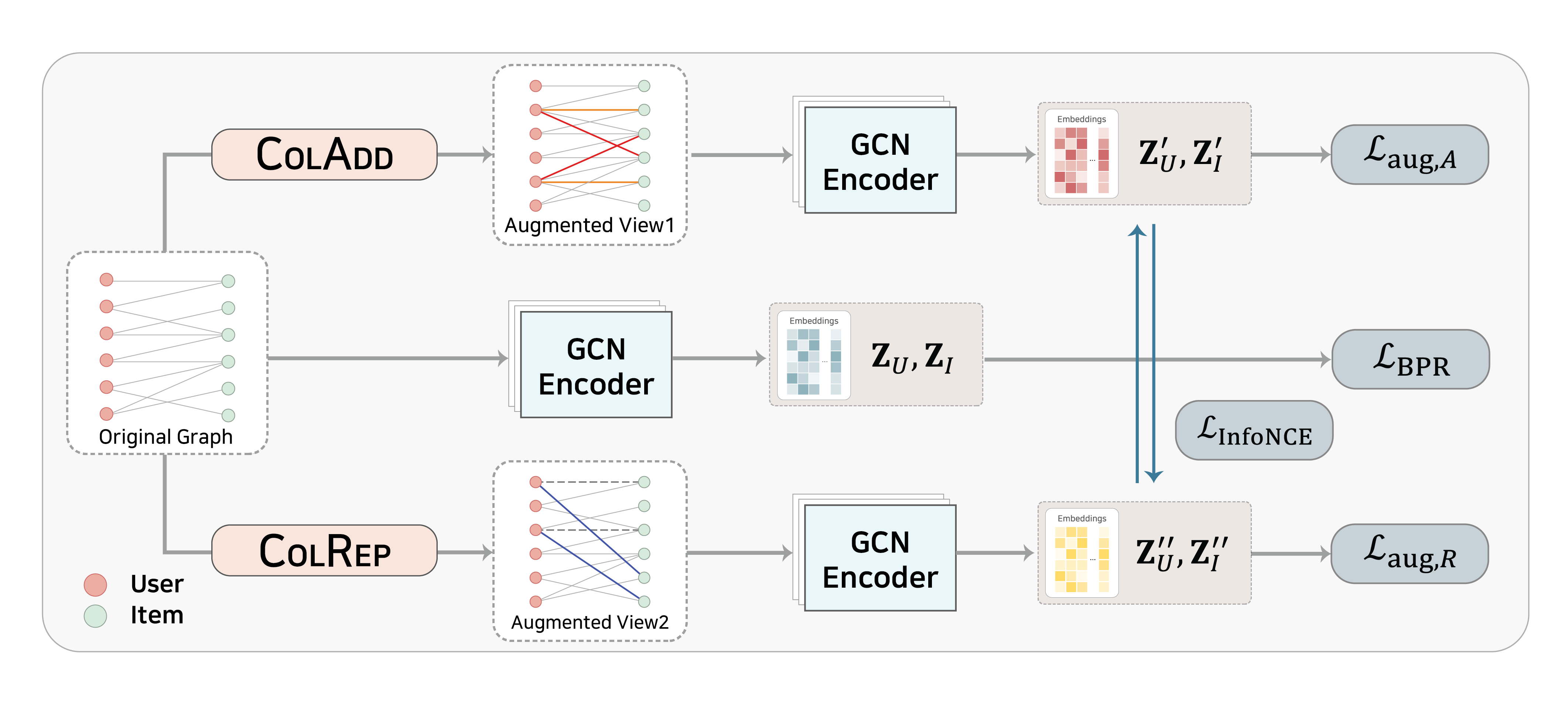}
    \vspace{-25pt}
    \caption{The overall framework of \method. At each training epoch, \moduleadd and \modulerep generate augmented views by inserting pseudo-interactions into the data, and these views are used for contrastive learning. The original graph's representations are used for the main recommendation task.}
    \vspace{-15pt}
    \label{fig:overall_framework}
\end{figure}

\begin{figure*}[t]
    \centering
    \vspace{-15pt}
   \includegraphics[width=\textwidth , trim=0.5cm 0cm 0.5cm 0, clip]{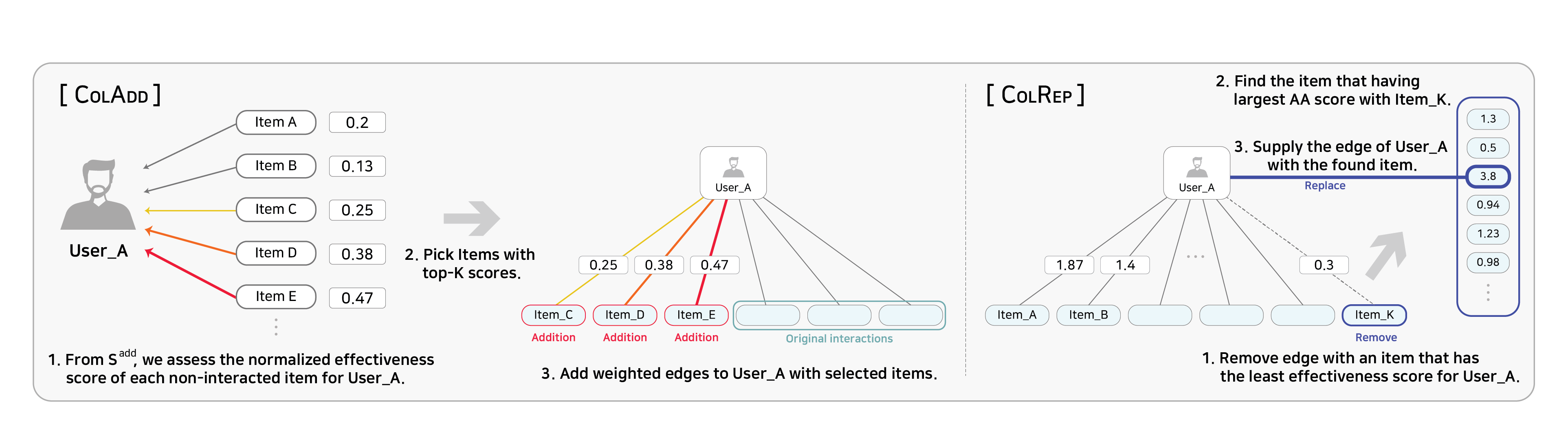}
    
    \vspace{-10pt}
    \caption{Visualization of the two \method augmentation methods: \moduleadd (left) and \modulerep (right). For \moduleadd, the top-$k$ items with the highest normalized effectiveness scores are sampled for each randomly selected user and then added through new weighted edges.
    On the other hand, \modulerep replaces the edges of selected users connected to the least effective items with new edges to the items that have the most similar behavioral features to the replaced items.}
    \vspace{-10pt}
    \label{fig:augmentation detail}
\end{figure*}
\section{Proposed Method} \label{sec3.proposed_method}



We propose \method, a framework for maximizing the use of collaborative information for effective augmentation.
We introduce two augmentation functions designed to preserve core interactions while providing meaningful signals. 



\subsection{Overview and Motivating Experiment}
\label{ssec:Motivating Experiment}

Figure \ref{fig:overall_framework} illustrates an overview of \method.
The two augmentation functions, \moduleadd and \modulerep, are used to create different augmented views at each training epoch.
The augmented and original graphs are passed through the GNN encoder to generate separate representations.
The representations from the augmented graphs are used for CL, while the representation from the original graph is used for the main recommendation loss.

\moduleaddlong (\moduleadd) adds pseudo-interactions between users and items that are highly relevant to each user.
By leveraging collaborative information actively, it identifies items that are likely to enhance each user's representation.
\modulereplong (\modulerep) replaces the least effective edge in a user's interaction history with a pseudo-interaction.
The pseudo-interaction connects the user to the item most similar to that being replaced.
Compared to previous methods that remove signals considered as noise, our approach focuses on providing additional meaningful signals for the CL process while minimizing the loss of core interactions and generating diverse augmented views.



\textbf{Motivating experiment.}
Before introducing the details of our method, we provide an intuitive motivation through a toy experiment.
As mentioned earlier, many existing CL-based recommendation approaches propose (learnable) denoising techniques, which remove noisy interactions from the original graph to create better contrastive views.
However, we argue that removing observed edges—based solely on implicit interactions—always carries the risk of discarding core signals essential for modeling user preferences. Such removal may introduce harmful biases during CL.

For the experiment, we use the Yelp and LastFM datasets, which are also used in our main experiments.
We first train a LightGCN encoder \cite{10.1145/3397271.3401063} to obtain robust user and item embeddings.
Then, using the same encoder initialized with these pre-trained embeddings, we generate final representations for users and items on graphs augmented by two existing methods and our \moduleadd and \modulerep.
Our goal is to assess whether each augmented graph preserves the semantic information necessary for accurate recommendations.

For the baselines, we consider (1) random interaction removal, following SGL \cite{10.1145/3404835.3462862}, where $10\%$ of the interactions are removed; and (2) a learnable denoising network from AdaGCL \cite{10.1145/3580305.3599768}, pre-trained with the best hyperparameters we identified. For fairness, we also apply our augmentation methods using the same perturbation ratio as in SGL.
We repeat the experiment five times with different random seeds and report the mean and standard deviation to account for the variability introduced by random augmentation.

Figure~\ref{fig:Validity of Denoising} shows the result. Unlike our augmentation functions, graphs from baselines consistently degrade performance compared to the original graph.
Notably, the complex learnable denoiser performs worse than random denoising and exhibits higher variance. We attribute this to the denoising methods relying on implicit feedback, which lacks explicit labels or reliable indicators of user preference. This makes it difficult to distinguish meaningful from spurious interactions, increasing the risk of removing important signals and leading to higher sensitivity to training data. These results highlight the risk of harmful bias introduced by augmented graphs generated through model-guided denoising in CL.

\subsection{Collaborative Edge Addition}
\label{ssec:addition}

As shown in Figure \ref{fig:augmentation detail}, \moduleadd creates an augmented view of the given graph by adding \emph{pseudo-interactions} between users and items, rather than removing ``noisy information'' which is difficult to define or quantify due to the nature of interaction data.
Inspired by the foundational idea of collaborative filtering (CF), we propose two hypotheses for generating effective pseudo-interactions.

\begin{proposition}[item-based pseudo-interactions]
An item $i$ is an effective candidate for a user $u$ if $i$ is similar to the items that $u$ has already interacted with.
\label{prop:item-add}
\end{proposition}

\begin{proposition}[user-based pseudo-interactions]
An item $i$ is an effective candidate for a user $u$ if $i$ has interactions with many other users having high interaction similarity with $u$.
\label{prop:user-add}
\end{proposition}

Given a user $u$, the two hypotheses differ in how they define the effectiveness of an item $i$ for a user $u$ who has no previous interactions with $i$.
In Proposition~\ref{prop:item-add}, the similarity is measured between $i$ and $N_I(u)$, which is the set of items that $u$ has interacted with, while in Proposition~\ref{prop:user-add}, the effectiveness is determined by the strength of similarities between $u$ and $N_U(i)$, which is the set of users who have interacted with $i$.
That is, they are different ways to utilize the idea of CF for creating augmented views.


To design a function that generates pseudo-interactions following each proposition, we first measure the similarity between nodes of the same type: \emph{user-user} and \emph{item-item} and store each in a matrix of size ${m\times m}$ and ${n\times n}$, respectively.
We use the Adamic-Adar $(\mathrm{AA})$ score~\cite{ADAMIC2003211}, which effectively captures the strength of relationships based on the neighbors:
\begin{equation}
\label{eq:AA}
    \mathrm{AA}(u,v) =  {\textstyle \sum_{k \in N(u) \cap N(v)}} \frac{1}{\log(d(k))},
\end{equation}
where $N(\cdot)$ denotes the set of 1-hop neighbors and $d(\cdot)$ is the node degree, representing their count.

Then, we propose two types of effectiveness scores, which are stored in $\mathbf{S}^{\mathrm{user}}, \mathbf{S}^{\mathrm{item}}\in\mathbb{R}^{m\times n}$, respectively:
\begin{align}
\label{eq:effective score_calculation}
    & s_{u,i}^\mathrm{user} = w_{i} \cdot {\textstyle \sum_{v \in N_U(i)}} \mathrm{AA}(u,v), \\
    & s_{u,i}^\mathrm{item} = w_{u} \cdot {\textstyle \sum_{j \in N_I(u)}} \mathrm{AA}(i,j),
\end{align}
where the weights $w_{i}= 1/d(i)$ and $w_{u} = 1/d(u)$ are introduced to balance the scale of scores across different users and items. 
The user-based effectiveness score $s_{u,i}^\mathrm{user}$ is based on the behavioral similarity between user $u$ and the direct neighbors of item $i$, while the item-based effectiveness score $s_{u,i}^\mathrm{item}$ is based on the similarity between item $i$ and the direct neighbors of user $u$.

Instead of using the raw scores, we propose to normalize them for each user, because we care about the relative importance of items to each user, not their absolute importance.
\begin{equation}
\label{eq:normalized effective score}
\tilde{s}_{u,i}^\mathrm{user} = \frac{s_{u,i}^\mathrm{user} - \min(\mathbf{S}_{u,*}^\mathrm{user})}{\max(\mathbf{S}_{u,*}^\mathrm{user}) - \min(\mathbf{S}_{u,*}^\mathrm{user})}.
\end{equation}
The same normalization is done also for $s_{u,i}^{\mathrm{item}}$.
Once the score matrices are normalized, we remove elements that correspond to the existing interactions to avoid redundancy:
\begin{equation}
\mathbf{S}^\mathrm{add} = \tilde{\mathbf{S}} \odot (\mathbf{1} - \mathbf{R}),
\end{equation}
where $\tilde{\mathbf{S}}$ is the normalized matrix of $\mathbf{S}$, which commonly denotes either $\mathbf{S}^{\mathrm{user}}$ or $\mathbf{S}^{\mathrm{item}}$. For user-based and item-based $\mathbf{S^{\mathrm{add}}}$  $\in \mathbb{R}^{m \times n}$, we randomly select one of these score matrices at each epoch to diversify the pools of generated views.  

After selecting the score criterion, we randomly sample a set of users by the ratio of $\rho_{\mathrm{add}}$, which is a hyperparameter.
We retain the top-$k$ items with the highest scores in $\mathbf{S}^\mathrm{add}$ for each sampled user, aiming to add the items most effective for modeling the user's behavioral characteristics to pseudo-interactions.
This is achieved by creating a mask matrix $\mathbf{M}^{\mathrm{add}}$, which is multiplied with $\mathbf{S}^\mathrm{add}$ as follows:
\begin{equation}
\mathbf{R}^{\mathrm{add}} = \mathbf{M}^\mathrm{add} \odot \mathbf{S}^\mathrm{add}.
\end{equation}
The matrix $\mathbf{R}^{\mathrm{add}}$ is then added to the original matrix $\mathbf{R}$, forming $\mathbf{R}^\mathrm{aug}_A$  that incorporates both existing and new interactions:
\begin{equation}
\mathbf{R}^\mathrm{aug}_A = \mathbf{R} + \mathbf{R}^{\mathrm{add}}.
\end{equation}

From the updated relation matrix, we generate a new adjacency matrix that reflects the augmented graph structure. This new adjacency matrix is used as input for the GCN encoder, which generates updated node representations that capture both the original and newly introduced collaborative signals. By integrating pseudo-interactions into the graph, the GCN encoder is able to learn more robust and comprehensive representations of users and items.

\subsection{Collaborative Edge Replacement}
\label{ssec:replacement}


We propose \modulerep to create a contrasting view to \moduleadd.
Our goal is to provide diverse variations of the original graph while minimizing the risk of losing core interactions.
To formalize this, we propose the following:
\begin{proposition}[unrelated items]
    Given a user $u$, an item $i$ is likely unrelated to $u$'s actual preference if $i$ has the lowest effectiveness score among the items $u$ has interacted with.
\label{prop:colrep}
\end{proposition}

To minimize the loss of important interactions, we aim to perturb only the interactions with items that are least relevant to the user's preferences during the augmentation process.
Based on the above Proposition~\ref{prop:colrep}, \modulerep perturbs only the interactions with items that are furthest from the user's core preferences, as determined by the effectiveness score we previously proposed.

As shown in Figure \ref{fig:augmentation detail}, \modulerep uses the same effectiveness and $\mathrm{AA}$ scores as in \moduleadd.
Similar to \moduleadd, one of the two pre-computed effectiveness score matrices is randomly selected at each training epoch.
After the score criterion is chosen, we randomly sample a set of users, with the proportion controlled by a hyperparameter $\rho_{\mathrm{rep}}$, to remove one of their interactions. For each sampled user, all interacted items are masked except for the one with the lowest effectiveness score, using a mask matrix $\mathbf{M}^{\mathrm{remove}}$, resulting in the construction of a new matrix $\mathbf{R}^{\mathrm{remove}}$:
\begin{equation} \mathbf{R}^{\mathrm{remove}} = \mathbf{M}^{\mathrm{remove}} \odot \mathbf{R}. \end{equation}


We then replace the removed edges with pseudo-interactions between each user and the item most similar to the removed item based on the pre-computed similarity score.
To implement this, we mask all non-interacted items except for one with the highest similarity to the removed item using $\mathbf{M}^{\mathrm{replace}}$: 
\begin{equation}
\mathbf{R}^{\mathrm{replace}} = \mathbf{M}^{\mathrm{replace}} \odot (\mathbf{1} - \mathbf{R}).
\end{equation}
That is, for each removed item $i$ for user $u$, we retain only the item $j$ that has the highest similarity to $i$ from non-interacted items, expressed as $j=\arg\max_{k\in I\setminus N(u)}\mathrm{AA}(i,k)$.
We set \smash{$m_{u,j}^{\mathrm{replace}}$} to $1$, while it remains $0$ for all other items.

We get the augmented matrix $\mathbf{R}^{\mathrm{aug}}_R$ by subtracting $\mathbf{R}^{\mathrm{remove}}$, which accounts for the removed edges, from $\mathbf{R}$, and adding the pseudo-interactions represented by $\mathbf{R}^{\mathrm{replace}}$ to it:
\begin{equation}
    \mathbf{R}^{\mathrm{aug}}_R = \mathbf{R} - \mathbf{R}^{\mathrm{remove}} + \mathbf{R}^{\mathrm{replace}}.
\end{equation}
With $\mathbf{R}^{\mathrm{aug}}_R$, we generate a new adjacency matrix for the graph, which is then used to create an augmented view. Through this process, we obtain safer and more diverse views.

\begin{figure*}[!t]
    \centering
    \vspace{-15pt}
    \includegraphics[width=\textwidth , trim=0.5cm 0 0.5cm 0, clip]{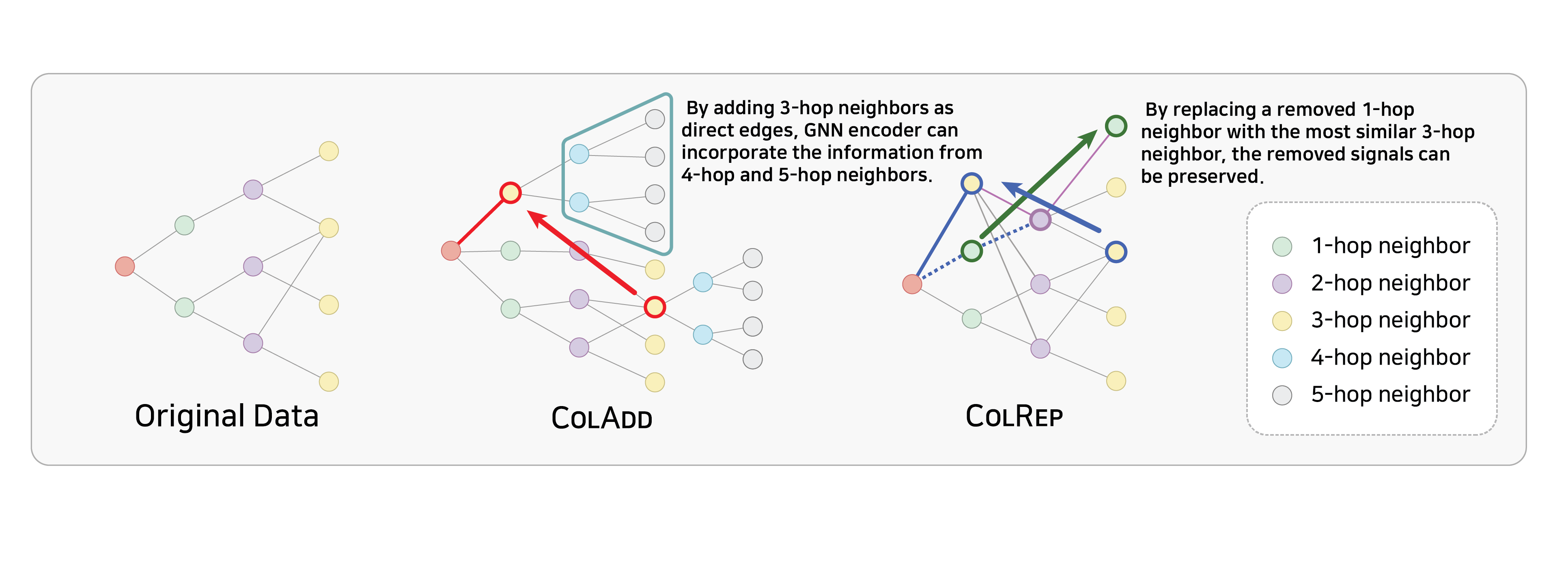}
    \vspace{-45pt}
    \caption{
        Illustration on how our augmentation methods reflect multi-hop collaborative signals and preserve core node features. In \moduleadd (center), original interactions remain untouched, while pseudo-interactions with 3-hop items are added, allowing up to 5-hop collaborative signals to be incorporated into the representation. In \modulerep (right), the least effective edge is replaced with an edge connected to a 3-hop item, partially preserving the 2-hop signals lost due to edge removal, while retaining signals from the removed edge. This process also incorporates multi-hop collaborative signals, similar to \moduleadd.
    }
    \vspace{-15pt}
    \label{fig:method_analysis_for_GNN_supplement}
\end{figure*}

\subsection{Training Objective}
\label{ssec:objective}

For effective training, we propose a multi-task objective that optimizes both the downstream and CL tasks:
\begin{equation}
    \mathcal{L}_{\text{total}} = \mathcal{L}_{\text{BPR}} + \lambda_{1}\mathcal{L}_{\text{InfoNCE}} + 
    \lambda_{2}\mathcal{L}_{\text{reg}} +
    \lambda_{3}\lVert\theta\rVert^{2}_{2},
\end{equation}
where $\mathcal{L}_{\text{BPR}}$ is the BPR loss \cite{rendle2012bprbayesianpersonalizedranking} for the main recommendation task, while $\mathcal{L}_{\text{InfoNCE}}$ is the CL loss which utilizes the augmented views generated from our augmentation methods.
We adopt the widely used InfoNCE loss \cite{oord2019representationlearningcontrastivepredictive} defined as follows:
\begin{equation}
     \mathcal{L}_{\text{InfoNCE}}^{\mathrm{user}} = -{\sum_{u \in U} \log \frac{\exp(\text{sim}(\mathbf{z}_{u}^{'}, \mathbf{z}_{u}^{''}) / \tau)}{\sum_{v\in U} \exp(\text{sim}(\mathbf{z}_{u}^{'}, \mathbf{z}_{v}^{''}) / \tau)}},
\end{equation}
where $\text{sim}(\cdot)$ measures the cosine similarity between node representations from the different views, and $\tau$ is a hyperparameter.
We compute the contrastive
loss $\mathcal{L}_{\text{InfoNCE}}^{\mathrm{item}}$ for the item side in a similar way.
The final CL loss is then obtained as the sum of two sides: $\mathcal{L}_{\text{InfoNCE}} =\mathcal{L}_{\text{InfoNCE}}^{\mathrm{user}}+\mathcal{L}_{\text{InfoNCE}}^{\mathrm{item}}$.

We further use a regularizer $\mathcal{L}_{\text{reg}}$  to ensure that the representations of augmented views remain well-aligned with the downstream task while enriching the collaborative information.
To this end, we apply the binary cross-entropy loss, which is widely used in recommender systems \cite{10.1145/3038912.3052569, 10.1145/2168752.2168771}:
\begin{multline} 
\mathcal{L}_{\text{reg}} = -\sum_{*\in \{A, R\}}\sum_{(u,i) \in E} y_{u,i} \log \hat{y}_{u,i}^{\text{aug,}*} \\ + (1 - y_{u,i}) \log (1 - \hat{y}_{u,i}^{\text{aug,}*}),
\end{multline}%
\noindent where \smash{$\hat{y}_{u,i}^{\text{aug,}*}$} represents predictions based on representations from the augmented graph generated by $\mathbf{R}_A^{\mathrm{aug}}$ and $\mathbf{R}_R^{\mathrm{aug}}$.

\subsection{Complexity Analysis}
\label{ssec:time_complextiy_analysis}
\begin{table}
\small
\centering
\caption{Time complexity of CL-based GCF methods.}
\vspace{-5pt}
\label{Table: Time complexity}
\begin{threeparttable}
\resizebox{\linewidth}{!}{
\begin{tabular}{l|lll}
\toprule
\textbf{Method} & \textbf{Pre-calculation} & \textbf{Augmentation} & \textbf{Graph Convolution} \\
\midrule
LightGCN  & - & - & $O(2|E|LTd)$ \\
SGL$^{*}$  & - & $O(2\rho |E|T)$ & $O((2 + 4\rho)|E|LTd)$ \\
NCL  & $O(2|E|KT_kd)$ & $O((m+n)Td)$ & $O(2|E|LTd)$ \\
HCCF$^{*}$  & - & - & $O(2(|E| + H(m+n))LTd)$ \\
LightGCL$^{*}$  & $O(2q|E|)$ & - & $O(2(|E| + q(m+n))LTd)$ \\
DCCF$^{*}$ & - & $O(2|E|LTd)$ & $O(2(|E|+K(m+n))LTd)$ \\
AdaGCL  & - & $O(|E|LTd^2)$ & $O(6|E|LTd)$ \\
\midrule
SCAR$^{\dagger}$ & $O(\alpha)$ & $O((2|E|+k\rho_{\mathrm{add}}m)T)$ & $O((6|E|+2k\rho_{\text{add}}m)LTd)$\\
\bottomrule
\end{tabular}
}
\begin{tablenotes}
\footnotesize
\setlength{\labelsep}{0pt}
\setlength{\itemindent}{-10pt}
\item ${*}$ Complexities are taken from their original papers.
\item${\dagger}$ \parbox[t]{\columnwidth}{
$\alpha = \mathrm{NNZ}(\mathbf{S}^{\mathrm{add}}) = \mathrm{NNZ}(\mathbf{R}\mathbf{R}^T\mathbf{R}) - \mathrm{NNZ}(\mathbf{R})$, where $\mathrm{NNZ}$ counts nonzero entries.
}
\end{tablenotes}
\end{threeparttable}
\vspace{-15pt}
\end{table}

\method is highly scalable to large data.
In Table \ref{Table: Time complexity}, we analyze the time complexity of \method and compare it to other CL-based GCF baselines. The training time for these methods can be divided into four main components: pre-calculation, data augmentation, graph convolution, and contrasting augmented views.
We report only the first three in the table, since the time for contrasting augmented views is the same across all methods and linear with batch size.

\noindent\textbf{Pre-calculation.}
The time complexity for the pre-calculation of \method is proportional to the number of non-zero entries in $\mathbf{S}^{\mathrm{add}}$, which we denote as $\alpha$ in the table.
$\mathbf{S}^{\mathrm{add}}$ is highly sparse in most cases, due to the sparsity and the long-tail distribution of interactions.
As supportive evidence, we report the actual pre-computation time on the Amazon dataset \cite{10.1145/3616855.3635814}, which contains approximately 1 million interactions.
\method takes $0.4$ seconds for computing the AA scores, $143.8$ for \moduleadd, and $76.0$ for \modulerep.
This is negligible overhead, compared to the training phase where each epoch takes around 36 seconds.
Moreover, the pre-computation step is executed only once and its output can be reused across multiple runs.

\noindent\textbf{Data augmentation.} The time complexity of the data augmentation phase in \method is $O((2|E|+k\rho_{\mathrm{add}}m)T)$, where $|E|$ is the number of interactions, $\rho_{\mathrm{add}}$ represents the proportion of users selected for \moduleadd, $k$ is the number of items added for each sampled user, and $T$ is the number of training epochs.
Specifically, generating the augmented relation matrix $\mathbf{R}^{\mathrm{aug}}_A$ for \moduleadd requires $O(|E|+k\rho_{\mathrm{add}}m)$, while $\mathbf{R}^{\mathrm{aug}}_R$ for \modulerep requires $O(|E|)$.

\noindent\textbf{Graph convolution and contrasting augmented views.} The time complexity of the graph convolution step in \method is $O((6|E|+2k\rho_{\text{add}}m)LTd)$, where $d$ is the embedding dimension. For contrasting augmented views, the complexity across all baselines is $O(Bd+ BMd)$ per batch, where $B$ is the batch size and $M$ is the number of users and items in the batch.

\subsection{Discussion}
We further discuss why our approaches are effective and can outperform previous augmentation functions for GCF.

\noindent\textbf{Capturing multi-hop signals.} \method generates views that help the GNN encoder leverage multi-hop collaborative signals beyond the number of layers $L$ through contrastive learning, enabling it to create more effective representations. GNN encoders typically face challenges like oversmoothing and oversquashing \cite{Li_Han_Wu_2018, Chen_Lin_Li_Li_Zhou_Sun_2020}, which limit the number of GNN layers to be used.
\method addresses these issues by generating 3-hop pseudo-interactions to the graph.
Given a target user $u$, items selected for its pseudo-interactions are always 3-hop neighbors of $u$, since they are direct neighbors of the 2-hop users of $u$ but have not directly interacted with $u$.

As illustrated in Figure \ref{fig:method_analysis_for_GNN_supplement}, \method enhances the aggregation process by promoting pseudo-interactions between the ego node and its 3-hop neighbors, treating them as 1-hop neighbors for GNN aggregation. This allows us to incorporate collaborative signals from farther nodes, reaching up to $L+2$ hops, which would otherwise be excluded due to layer limitations in standard GNNs. This feature of \method allows the representations, enriched with multi-hop collaborative signals, to be effectively utilized in CL, significantly aiding the encoder in modeling user behavior more accurately. 
\begin{table}[ht]
\selectfont
\begin{center}
\caption{Dataset statistics and interaction sparsity levels.}
\label{Table: CombinedStatistics}
\vspace{-5pt}
\resizebox{\columnwidth}{!}{
\begin{tabular}{l|ccc|C{0.85cm}C{0.85cm}C{0.85cm}}
\toprule
\multirow{2.5}{*}{\textbf{Dataset}} 
& \multirow{2.5}{*}{\textbf{\# Users}} 
& \multirow{2.5}{*}{\textbf{\# Items}} 
& \multirow{2.5}{*}{\begin{tabular}[c]{@{}c@{}}\textbf{\# Interactions}\\\textbf{\small(Density)}\end{tabular}} 
& \multicolumn{3}{c}{\textbf{User-wise \# Interactions}} \\
\cmidrule(lr){5-7}
& & & & $\mathbf{\leq 5}$ & $\mathbf{\leq 10}$ & $\mathbf{> 10}$ \\
\midrule
Yelp & 42,712 & 26,822 & 182,357 ($1.6\times10^{-4}$)  & 78\% & 14\% & 8\% \\
Gowalla & 25,557 & 19,747 & 294,983 ($5.8\times10^{-4}$) & 39\% & 26\% & 35\% \\
Amazon & 76,469 & 83,761 & 966,680 ($1.5\times 10^{-4}$) &23\%&40\%&37\%\\
LastFM & 1,892 & 17,632 & 64,983 ($2.0\times 10^{-3}$) & 0.2\% & 0.3\% & 99\% \\
\bottomrule
\end{tabular}}
\end{center}
\vspace{-10pt}
\end{table}

\begin{table*}[thbp]
\selectfont
\begin{center}
\caption{
    Performance comparison between \method and state-of-the-art CL baselines on four datasets.
    Results are averaged over five random seeds.
    Across all datasets and metrics, \method achieves the highest average recommendation performance. The best results are in \textbf{bold}, and the second-best are \underline{underlined} (R = Recall, N = NDCG).
}
\label{Table:Model Comparison_baseline}
\vspace{-5pt}

\huge

\resizebox{\textwidth}{!}{
\begin{tabular}{l|cccc|cccc|cccc|cccc|c}
\toprule
\multirow{2}{*}{\textbf{Method}} & \multicolumn{4}{c|}{\textbf{Gowalla}} & \multicolumn{4}{c|}{\textbf{Yelp}} &\multicolumn{4}{c|}{\textbf{Amazon}}&\multicolumn{4}{c|}{\textbf{LastFM}}& Avg\\
        \cmidrule{2-17}
        & R@10 & R@20 & N@10 & N@20 
        & R@10 & R@20 & N@10 & N@20
        & R@10 & R@20 & N@10 & N@20 & R@10 & R@20 & N@10 & N@20 & Rank\\ 
        \midrule

LightGCN & .0905 & .1347 & .0772 & .0902  & .0383 & .0588 & .0232 & .0293 &.0354 & .0583& .0293& .0371  & .0927 & .1373 & .0702 & .1020& 9.00\\
\midrule
SGL & .1636 & .2374 & \underline{.1335} & .1555  & .0544 & .0846 & .0337 & .0426 &\underline{.0720} &\underline{.1060} &\underline{.0614} &\underline{.0726}  & \underline{.1724} & \underline{.2495} & \underline{.1524} & \underline{.1844}&\underline{2.75}  \\
NCL & .1477 & .2195 & .1186 & .1403  & .0522 & .0813 & .0328 & .0414 & .0570 & .0855 & .0486 & .0581& .1697 & .2473& .1506 & .1829 & 5.13 \\
HCCF & .1504 & .2260 & .1181 & .1412  & .0438 & .0693 & .0264 & .0340 & .0598 & .0935& .0485 & .0600  & .1660 & .2403 & .1441 & .1751 & 6.13 \\
LightGCL & .1568 & .2313  & .1257 & .1483  & .0529 & .0824& .0327 & .0414& .0454 & .0700 & .0379 & .0462 & .1519 & .2196 & .1352 & .1634 &6.56\\
DCCF & \textbf{.1691} & \textbf{.2462} & \textbf{.1364} & \textbf{.1597}  & .0552 & \underline{.0861} & .0347 & .0437& .0493 & .0734 & .0444 & .0520 & .1623 & .2346 &  .1428 & .1731 & 4.31 \\
AdaGCL & .1386 & .2078  & .1130 & .1339 & .0504 & .0787& .0308 & .0392 & .0471 & .0746 & .0387 & .0480  & .1673 & .2434 & .1484 & .1801 &6.69 \\
MixSGCL & .1576 & .2343 & .1278 & .1508 & \underline{.0563} & \textbf{.0880} & \underline{.0352} & \textbf{.0446}& .0669 & .1027 & .0589 & .0706  & .1716 & .2472 & .1509 & .1825 & 3.00 \\
\midrule
\textbf{SCAR (ours)} & \underline{.1689} & \underline{.2445} & \textbf{.1364} & \underline{.1591} & \textbf{.0566} & \textbf{.0880} & \textbf{.0353} & \underline{.0445} & \textbf{.0736}& \textbf{.1080}& \textbf{.0618}&\textbf{.0733} & \textbf{.1746} & \textbf{.2531} & \textbf{.1543} & \textbf{.1870} & \textbf{1.25}  \\
\bottomrule 
\end{tabular}
}
\vspace{-15pt}
\end{center}

\end{table*}

\noindent\textbf{Minimizing the loss of core interactions.}
\method is designed to minimize the risk of core interactions being excluded from the representation learning process. In \moduleadd, no existing interactions are lost; instead, we enhance the representations by adding meaningful collaborative signals to the existing ones. In \modulerep, while some edges are removed, we mitigate this by removing only the edge to the least effective item based on the effectiveness score. Additionally, by supplementing the lost interaction with an item that has the most similar behavior, we preserve the signals from 2-hop neighbors that would otherwise be lost. Moreover, as shown in Figure \ref{fig:method_analysis_for_GNN_supplement}, replacing these edges ensures that the removed item remains part of the 3-hop aggregation, guaranteeing its contribution is still reflected in the overall representation learning process.
As a result, both of our augmentation algorithms can create representations that minimize the loss of core information.
This significantly reduces the risk of introducing incorrect biases into the representations during the CL process.

\noindent\textbf{Interpretable augmentation.}
The two augmentation functions in \method are \emph{interpretable} as they allow clear analysis of structural signals being introduced or replaced during the training.
The effectiveness scores guiding these augmentations are computed from interaction similarities between users and items.
This enables direct reasoning about why a particular perturbation (e.g., addition or replacement) is applied to a specific pair of nodes.
Beyond transparency, these behavior-driven augmentations enable diagnostic analysis.
For instance, by examining the popularity of items (e.g., degree statistics) for each user, which is related to the reliability of the effectiveness score, one can better control the perturbation ratio.
Refer to Section \ref{ssec:case-study} for the case studies of \method on a real-world dataset for deeper insights on its interpretability.



\section{Experiments}
We run all experiments based on SSLRec \cite{10.1145/3616855.3635814}, a framework for self-supervised learning-based recommender systems, to enable fair, consistent and reproducible experiments.

\noindent\textbf{Datasets.}
We use four datasets in experiments: Gowalla \cite{cho2011friendship}, Yelp\footnote{https://www.yelp.com/dataset}, Amazon \cite{10.1145/3616855.3635814} and LastFM \cite{schifanella2010folks}. 
Gowalla contains user check-ins at various locations, with interactions represented as user-location pairs. 
Yelp contains user interactions in the form of reviews and ratings for various business venues.
Amazon contains implicit user feedback on items in the book category from the Amazon e-commerce platform.
LastFM is a music recommendation dataset from the LastFM music platform. 
For Gowalla, Yelp and Amazon provided in SSLRec,
we follow the data split given by the framework, which uses the 14:1:5 ratio for training, validation, and test sets.
For the externally imported LastFM, we split the data by the 7:2:1 ratio for training, validation, and test, following a previous work \cite{10.1145/3580305.3599768}.
Detailed data statistics are given in Table \ref{Table: CombinedStatistics}.


\noindent\textbf{Baselines.}
We compare \method with state-of-the-art CL-based GCF approaches, including popular GCF models for comprehensive results.
The details are given as follows. 
\begin{itemize}[left=0pt]
    \item \textbf{LightGCN} (2020) \cite{10.1145/3397271.3401063} uses a linearized form of GCN as its encoder to perform efficient, consistent recommendation.
    \item \textbf{SGL} (2021) \cite{10.1145/3404835.3462862} enhances GCF with CL by performing augmentations using perturbations such as random dropout.
    \item \textbf{NCL} (2022) \cite{10.1145/3485447.3512104} generates both structural and semantic neighbors for the nodes and leverages them in CL-based structural and prototype objectives.
    \item \textbf{HCCF} (2022) \cite{10.1145/3477495.3532058} generates global views using hypergraphs and contrasts them with local views.
    \item \textbf{LightGCL} (2023) \cite{cai2023lightgclsimpleeffectivegraph} decomposes the adjacency matrix and applies a low-rank approximation to generate CL views.
    \item \textbf{DCCF} (2023) \cite{10.1145/3539618.3591665} generates augmented data by disentangling user intents to create effective CL views.
    \item \textbf{AdaGCL} (2023) \cite{10.1145/3580305.3599768} introduces data-adaptive augmentation by incorporating graph generation and denoising modules.
    \item \textbf{MixSGCL} (2024) \cite{zhang2024mixedsupervisedgraphcontrastive} applies tuned node- and edge-level mixup-based \cite{zhang2018mixupempiricalriskminimization} augmentations for CL.
    
\end{itemize}

\noindent\textbf{Evaluation metrics.}
We adopt the all-ranking protocol, where all non-interacted items in the training data for each user are considered as candidates to evaluate accuracy for each test user \cite{zhao2020revisiting}.
We employ Recall@N and NDCG@N as evaluation metrics \cite{gunawardana2009survey, steck2013evaluation}, where we set $N=10,20$ and report the average from five random seeds for each combination of a dataset and a metric \cite{10.1145/3616855.3635814}.

\noindent\textbf{Hyperparameter settings.}
For a fair comparison, all models are initialized with the Xavier initialization \cite{glorot2010understanding} and optimized using Adam \cite{kingma2014adam} with a learning rate of $10^{-3}$.
The batch size is 4096 and the node embedding size is 32 for all experiments.
We employ early stopping with Recall@10 as the stopping criterion. 
We thoroughly search for the best hyperparameters for all baselines, using the same approach as for our method.


\subsection{Overall Performance}

Table \ref{Table:Model Comparison_baseline} compares \method and the baselines in overall performance.
\method ranks first on average across all datasets and metrics.
This highlights the effectiveness of the augmentation functions designed for \method.
By minimizing the risk of core interaction loss and ensuring the incorporation of diverse collaborative information, \method significantly improves the encoder's learning process.
The most notable performance improvements are observed on Yelp and Amazon, the sparsest datasets. This suggests that the pseudo-interactions generated by \method are particularly beneficial for improving the representations of nodes with sparse interactions. They help the encoder learn core information from augmented views enriched with collaborative signals, leading to better representations.

On the Gowalla dataset, DCCF outperforms \method. We believe this may be due to several factors. First, DCCF employs a GNN encoder with additional modules that encode multi-intent information at each layer, enhancing its capacity to model complex user behavior. Moreover, it applies three distinct augmentation strategies for contrastive learning, whereas most other methods rely on one or two. This richer training setup may have contributed to stronger representations on Gowalla. However, when considering the overall results across all four datasets, the effectiveness of learnable approaches like DCCF appears highly sensitive to dataset characteristics. In contrast, \method consistently delivers robust performance regardless of data distribution, highlighting a key advantage of our interpretable and data-efficient design.



\begin{figure}[t]
\centering
\includegraphics[width=\columnwidth]{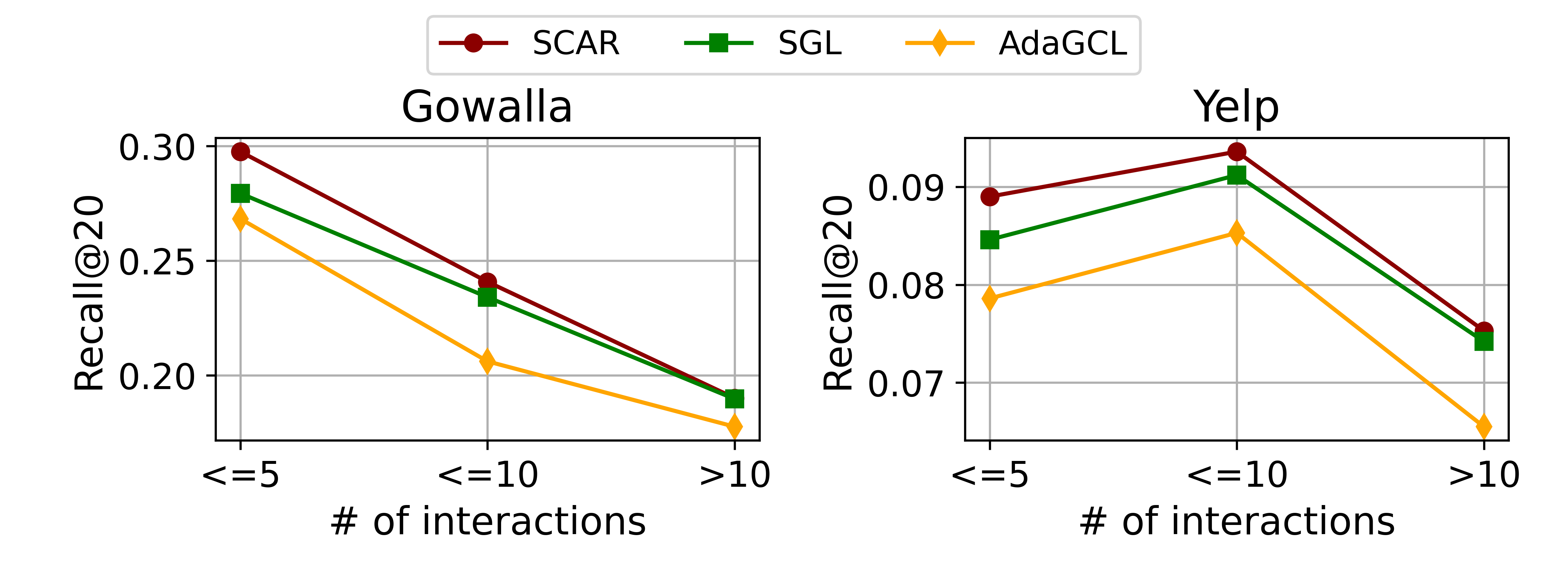}
\vspace{-25pt}
\caption{
    Performance across user groups with different interaction sparsity levels on Gowalla and Yelp.
    \method is generally effective and is particularly strong for sparse users.
}
\label{fig:Data Sparsity Comparision}
\vspace{-10pt}
\end{figure}
\begin{figure}
\centering
\includegraphics[width=\columnwidth]{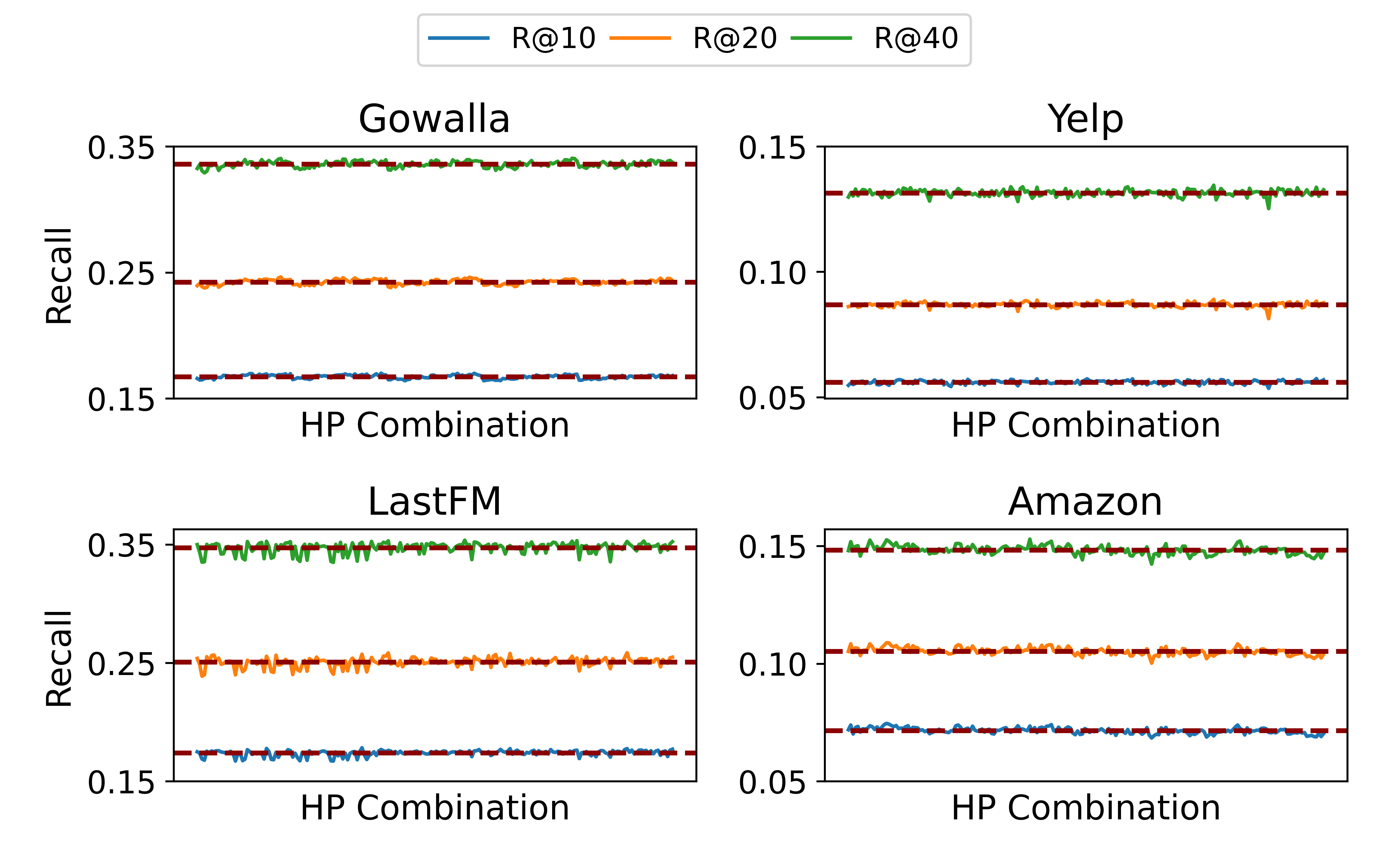}
\vspace{-20pt}
\caption{
    Robustness of \method to its four hyperparameters: ${(\rho_{\mathrm{add}},\rho_{\mathrm{rep}},k,\lambda_2)}$.
    \method remains highly stable, not deviating much from the average values denoted as the red lines.
}
\vspace{-15pt}
\label{fig:hyperparameter_analysis}
\end{figure}

\subsection{Effectiveness to Diverse Sparsity Levels}

We measure the accuracy of \method on user groups with different interaction sparsity levels in the Gowalla and Yelp datasets, comparing it with SGL and AdaGCL, the two competitive baselines.
Users are grouped based on their interaction count $i$: (a) $i \le 5$, (b) $5 < i \le 10$, and (c) $i > 10$.

Figure \ref{fig:Data Sparsity Comparision} illustrates the result.
\method demonstrates superior performance across all user groups in both datasets, and achieves the best improvement in the user group with the highest sparsity (no more than 5).
This suggests that our augmentation methods, which incorporate meaningful pseudo-interactions, effectively enrich collaborative signals in representations of sparse users.
\subsection{Robustness to Hyperparameter Settings}

Beyond the basic hyperparameters related to the GNN encoder and contrastive loss, \method introduces four additional hyperparameters: $(\rho_{\mathrm{add}},\rho_{\mathrm{rep}},k,\lambda_2)$.
We demonstrate that the performance of \method remains robust across different choices of these hyperparameters, which is important for the practical usability of a recommender system.
We conduct experiments over 200 hyperparameter combinations using the search ranges $\rho_{\mathrm{add}},\rho_{\mathrm{rep}} \in\{0.1, 0.2, 0.4, 0.6, 0.8\}$, $k \in \{3, 5, 7, 9\}$ and $\lambda_2 \in \{10^{-2}, 10^{-3}\}$, while keeping other hyperparameters fixed at their best settings.


As shown in Figure \ref{fig:hyperparameter_analysis}, \method demonstrates remarkable robustness to variations in the combination of its hyperparameters across all datasets.
This highlights the practical advantages of \method, simplifying the tuning process while ensuring stable and reliable performance across a wide range of settings.
Furthermore, we consider the average performance of \method shown as the red dotted lines as the random-hyperparameter version of it.
Then, we compare it with the other baselines as done in Table \ref{Table:Model Comparison_baseline}.
\method remains highly competitive even in this setting, achieving Recall@20 scores of 0.0869 on Yelp, 0.2424 on Gowalla, 0.1053 on Amazon, and 0.2506 on LastFM.
It ranks second, second, second, and first on the four datasets, respectively, out of the nine models.


\begin{table}
\selectfont
\centering
\caption{Comparison with masked autoencoder-based methods. \method outperforms them on 4 out of 6 settings.}
\label{Table:Method Comparison_MAE}
\vspace{-5pt}

\LARGE  

\resizebox{0.46\textwidth}{!}{
\begin{tabular}{l|cc|cc|cc}
\toprule
\multirow{2}{*}{\textbf{Method}} & \multicolumn{2}{c|}{\textbf{Gowalla}}&\multicolumn{2}{c|}{\textbf{Yelp}}&\multicolumn{2}{c}{\textbf{LastFM}} \\
\cmidrule{2-7}
 & R@20 & N@20 & R@20 & N@20 & R@20 & N@20 \\ 
\midrule

AutoCF & \textbf{.2561} & .1651 & .0844 & .0428 & .2277 & .1720 \\
Gformer & .2537 & \textbf{.1680} & .0734 & .0366 & .2474 & .1834\\
\midrule
\textbf{\method (ours)} & .2444 & .1590 & \textbf{.0874} & \textbf{.0444} &  \textbf{.2531} & \textbf{.1870} \\
\bottomrule
\end{tabular}
}

\end{table}
\begin{table}
\selectfont
\begin{center}
\caption{
    Ablation study on \method.
    We compare it with three variants, each removing one key module, and two additional variants that replace the AA score with other metrics. 
    Each module contributes to performance improvement, and \method shows strong robustness for similarity measures.
}
\label{Table:ablation study}
\vspace{-5pt}

    \resizebox{0.46\textwidth}{!}{
\begin{tabular}{l|cccc}
\toprule
\textbf{Method} & \textbf{R@20} & \textbf{N@20} & \textbf{R@40} & \textbf{N@40}  \\ 
\midrule
SCAR-wo-$\mathcal{L}_{\text{reg}}$ & .2460 & .1596 & .3405 & .1863 \\
SCAR-wo-\modulerep & .2371 & .1526 & .3331 & .1795 \\ 
SCAR-wo-\moduleadd & .2349 & .1512 & .3341 & .1790 \\ 
\midrule
\method-w-JC & .2438 & .1590 & .3378 & .1855 \\
\method-w-CN & .2439 & .1588 & .3359 & .1847 \\
\midrule
\textbf{SCAR (ours)} & \textbf{.2464} & \textbf{.1601} & \textbf{.3407} & \textbf{.1867} \\
\bottomrule 
\end{tabular}}
\vspace{-15pt}
\end{center}
\end{table}
\begin{figure*}[th]
    \centering
    \vspace{-15pt}
   \includegraphics[width=\textwidth, trim=1cm 0 1cm 0, clip]{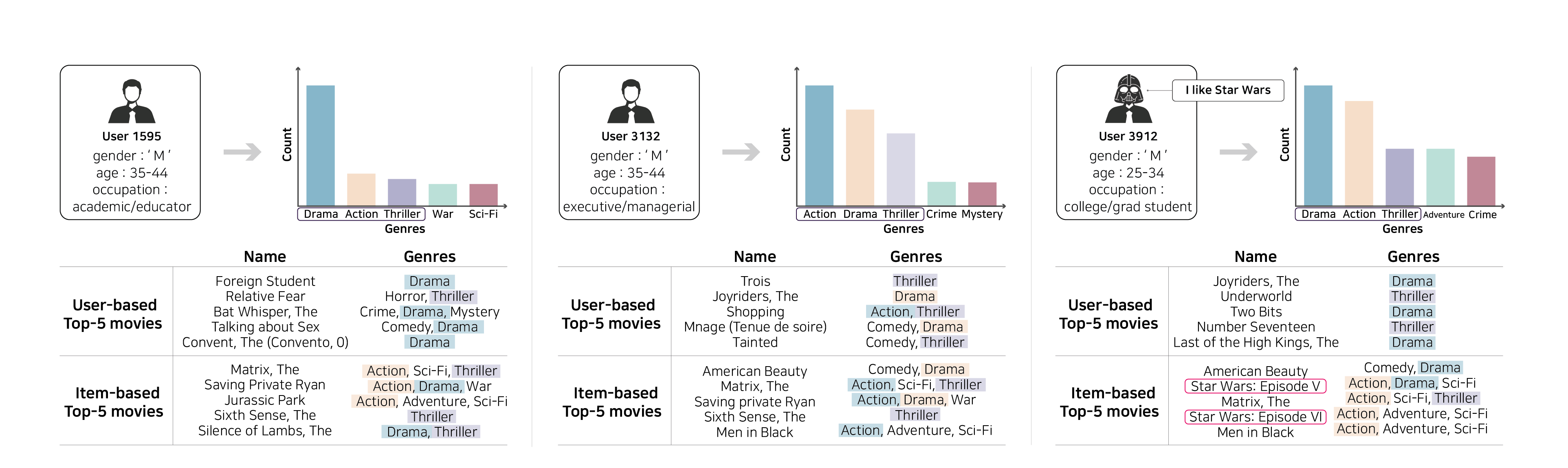}
    
    \vspace{-10pt}
    \caption{
        Case study on our effectiveness scores for capturing users' real preferences.
        We compare the top-5 movies ranked by our effectiveness scores with users' actual preferences.
        The results show that the pseudo-interactions created by our scores closely align with users' real preferences, helping the encoder capture users' preferences through contrastive learning.
    }
    \vspace{-15pt}
    \label{fig: Case study}
\end{figure*}
\subsection{Comparison with Masked Autoencoders}
\label{sec: Baseline Comparison with Mask AutoEncoder(MAE)}

To further validate the effectiveness of \method, we broaden our comparison to include recent non-CL-based SSL approaches.
Specifically, we evaluate \method against two state-of-the-art methods on three datasets, Gowalla, Yelp, and LastFM, that employ masked autoencoder (MAE)-based objectives.
AutoCF \cite{10.1145/3543507.3583336} uses an augmentation strategy that integrates structure-adaptive information within the MAE framework.
Gformer \cite{10.1145/3539618.3591723} incorporates invariant collaborative modules into the MAE paradigm, enhancing performance within a graph transformer architecture.



As shown in Table \ref{Table:Method Comparison_MAE}, \method achieves the best average rank (1.67) compared to GFormer (2.17) and AutoCF (2.17).
This superior performance is achieved with a simpler architecture and significantly lower computational complexity, avoiding reliance on the costly mechanisms in these baselines, such as graph transformer encoders with $O(|E|d^{2})$ complexity, which is significantly larger than our complexity $O(|E|Ld)$, or complex augmentation strategies. 
\subsection{Ablation Studies on Core Modules and Behavior Similarity} \label{sec: experiemtal 4.3}

We evaluate the effect of each key component in \method by removing it individually: \moduleadd, \modulerep, and $\mathcal{L}_{\text{reg}}$.
First, for \moduleadd and \modulerep, we replace each module with the embedding perturbation method used in SimGCL \cite{10.1145/3477495.3531937} (SCAR-wo-\moduleadd and SCAR-wo-\modulerep).
Second, to study the role of $\mathcal{L}_{\text{reg}}$, we compare the performance with and without this regularizer (SCAR-wo-$\mathcal{L}_{\text{reg}}$). Then, we compare the performance of each variant with the original version on the Gowalla dataset using Recall and NDCG at 20 and 40.

As shown in Table \ref{Table:ablation study}, removing either of the two augmentation modules, \moduleadd or \modulerep, results in a significant performance drop compared to the original model. This indicates that both augmentation modules play a crucial role in enhancing the effectiveness of CL. 
Similarly, removing $\mathcal{L}_{\text{reg}}$ makes a performance drop, indicating that regularizing the node embeddings by aligning augmented graph representations with the downstream task is crucial for maximizing the effectiveness of our augmentation.


Additionally, we evaluate the effect of the AA score in deriving effectiveness scores by comparing it with other similarity metrics: Jaccard similarity (JC) and common neighbors (CN).
\method demonstrates strong robustness to different metrics, showing superior performance with all three choices.



\subsection{Case Studies}\label{ssec:case-study}
We demonstrate the utility of the user-based and item-based effectiveness scores for generating pseudo-interactions through case studies.
To evaluate how well these scores align with actual user behaviors, we use the MovieLens dataset \cite{10.1145/2827872}, which is appropriate for deeper analysis since it contains user ratings for movies along with detailed information about both the movies and users.

We first split the dataset into training and test sets.
Then, we sample several users and calculate the normalized effectiveness scores for all movies using both user-based and item-based methods based on their interactions in the training data.
We then analyze the correlation between the top-5 movies selected based on the calculated scores (i.e., pseudo-interactions) and the users' actual preferences, which are derived by extracting the genres of the top-5 rated movies a user has interacted with and counting the frequency of each genre.
The results in Figure \ref{fig: Case study} demonstrate that our effectiveness scores generate meaningful pseudo-interactions for augmented graphs.
It presents the genre distribution of users’ highest-rated movies along with the top-5 selected movies based on calculated effectiveness scores. For the sampled users—User 1595, User 3132, and User 3912—the results indicate that the selected movies align well with their most preferred genres.

A closer look reveals that, despite User 1595’s biased genre preference, the movies selected through our approach reflect the user’s interests in a well-balanced manner. For User 3132, who rated only nine movies, the generated pseudo-interactions include items highly relevant to their preferences. In the case of User 3912, who gave a high rating to Star Wars IV, the pseudo-interactions include Star Wars V and Star Wars VI, both receiving high effectiveness scores. These cases demonstrate that our method effectively captures the behavioral patterns of users with diverse characteristics, including sparse preferences and niche interests. From this perspective, the information provided by our augmentation strategy significantly aids the encoder in modeling user preferences during CL.

\section{Conclusion}
\label{sec:conclusion}
In this paper, we propose \method, a novel framework that leverages collaborative information in user-item interactions to generate augmented data for contrastive learning (CL). We introduce two augmentation functions, \moduleadd and \modulerep, which generate views enriched with collaborative signals while minimizing the risk of losing core interactions. These functions enable diverse and effective views, allowing the encoder to capture core behavioral patterns and produce more accurate representations. Extensive experiments show that \method consistently outperforms existing CL-based methods and recent non-CL-based methods, maintaining superiority across varying levels of sparsity. For future work, we plan to explore strategies that better condense and utilize collaborative signals, aiming to develop models that are more effective and generalizable across diverse recommendation settings.

\section*{Acknowledgments}
{
\small
This work was supported by the National Research Foundation of Korea (NRF) grant funded by the Korea government (MSIT) (RS-2024-00341425 and RS-2024-00406985).
This work was supported by BK21 FOUR (Connected AI Education \& Research Program for Industry and Society Innovation, KAIST EE, No. 4120200113769).
Jaemin Yoo is the corresponding author.
}



\bibliographystyle{IEEEtran}
\bibliography{references}

\end{document}